\documentclass[preprint]{aastex}
\pdfoutput=1
\usepackage{amsmath}
\usepackage{natbib}
\usepackage{multirow}
\usepackage{graphicx}
\begin{document}


\title{Type Ia Supernova Intrinsic Magnitude Dispersion and the Fitting of Cosmological Parameters}

\author{A. G. Kim}
\affil{Physics Division, Lawrence Berkeley National Laboratory, Berkeley, CA 94720}
\email{agkim@lbl.gov}

\begin{abstract}
I present an analysis for fitting cosmological parameters from a Hubble Diagram of a standard candle with unknown intrinsic
magnitude dispersion.  The dispersion is determined from the data themselves, simultaneously with
the cosmological parameters.  This contrasts with the strategies used to date.  The advantages of the presented analysis
are that it is done in a single fit (it is not iterative), it provides a statistically founded and unbiased estimate of the intrinsic dispersion,
and its cosmological-parameter uncertainties account for the intrinsic dispersion uncertainty.  Applied to Type Ia supernovae,
my strategy provides a statistical measure to test for sub-types and assess the significance of any magnitude corrections applied to the calibrated candle.
Parameter bias and differences between likelihood distributions produced by the presented and currently-used fitters are negligibly small  for existing and projected supernova
data sets. 
\end{abstract}

\keywords{Supernovae: Data Analysis and Techniques}

\section{Introduction}

The homogeneous nature of Type Ia supernovae (SNe Ia) makes them a popular tool for measuring cosmological
distances.  After empirical corrections based on light curve shape, color, and spectral features,
the absolute magnitude (or distance modulus) of a supernova can be determined to
$\sim 0.12$ mag \citep{2007A&A...466...11G, 2007ApJ...659..122J,2008ApJ...681..482C,2009A&A...500L..17B}.
SNe Ia have been used to successfully measure the expansion rate of the universe \citep[the Hubble Constant;][]{2001ApJ...553...47F,2009ApJ...699..539R},
discover its accelerated expansion \citep{1998AJ....116.1009R,1999ApJ...517..565P},
and measure the properties of the dark energy responsible for that acceleration \citep{2009ApJ...700.1097H,2010ApJ...716..712A}.

The small  scatter in the peak brightness of SN Ia luminosities is inferred from the small residuals in their
Hubble Diagrams   \citep{1968AJ.....73.1021K}; the intrinsic supernova magnitude dispersion is measured from differences between observed magnitudes
and those predicted by the cosmological model, e.g.\ the linear Hubble law for low redshift.  Although there are theoretical
explanations for this dispersion including intrinsic progenitor properties, circumstellar dust, and viewing angle \citep[see e.g.][]{2006ApJ...649..939K,
2008ApJ...686L.103G, 2003ApJ...591.1110W}, in practice the amount of dispersion is determined empirically from the data themselves.

The luminosity dispersions of supernova subsets are statistics that can be used to
compare and identify SN~Ia subclasses.
The prevailing belief is that the intrinsic luminosity of an individual supernova, including line-of-sight effects,
is encoded non-trivially within a finite set of physical and geometric parameters.  The ``intrinsic'' dispersion arises from
our lack of observational access to all those parameters and incomplete knowledge of how to exploit those that are available.
It is possible that SN Ia subclasses with different average luminosities are responsible for some of
the intrinsic dispersion seen in current data.   Correlations between supernova light curves and spectral features \citep{2005ApJ...623.1011B,
2009A&A...500L..17B, 2009ApJ...699L.139W, 2010arXiv1011.4517F} and host galaxy
\citep{2010MNRAS.406..782S,2010ApJ...722..566L} give evidence that SNe Ia need to be modeled in finer detail using an expanded suite of data.
Likelihood surfaces of intrinsic dispersion for supernova subsets provide a statistical measure to test whether data are best described by a single intrinsic
dispersion. 

This paper presents the methodology for simultaneously fitting for the intrinsic dispersion and the cosmological parameters that specify the
dynamics of the cosmic expansion.
Although I present straightforward textbook likelihood analysis, it has yet to be applied on supernova-cosmology data.
My approach contrasts with that of \citet{2010arXiv1006.2141S}, who suggest using Monte Carlo analysis of statistics that are insensitive to  the intrinsic dispersion.
This paper is organized as follows: \S\ref{likelihood:sec} presents the likelihood
equation and contrasts it with the commonly used method.  Results of simulations are given in \S\ref{simulation:sec} that show the
quantitative differences between the results
of the two analyses.  I finish with conclusions in \S\ref{conclusions:sec}.

\section{The Likelihood}
\label{likelihood:sec}
Given a set of measured quantities $\mu_i$ with covariance $\mathbf{C}$ at known points $z_i$ and a model  $F(z_i;\boldsymbol{\theta})$ for the corresponding true values parameterized by
$\boldsymbol{\theta}$, the Gaussian likelihood $L$ can be expressed as
\begin{align}
\mathcal{L} &\equiv   -2\ln{L} \nonumber\\
&= \ln{\det \mathbf{C}} + (\boldsymbol{\mu}-F(\mathbf{z};\boldsymbol{\theta}))^T \mathbf{C}^{-1} (\boldsymbol{\mu}-F(\mathbf{z};\boldsymbol{\theta}))
\label{likelihood:eqn}
\end{align}
neglecting the irrelevant $2\pi$ term.
For a supernova-cosmology analysis $\mu_i$ and $z_i$ correspond to the estimators for the distance modulus and redshift of supernova $i$.
The function $F$ is the theoretical prediction for the distance modulus as a function of redshift and a set of cosmological parameters, e.g.\
$\Omega_M$, $\Omega_{DE}$, and $w$.

The covariance matrix gives the deviation of all possible measurements for all possible supernovae
from the mean distance moduli at the given redshifts; $\mathbf{C}$ not only has
a contribution from measurement uncertainty in supernova magnitudes, $\mathbf{C}_{m}$, but also from the fact that supernovae are drawn from a population
with intrinsic magnitude dispersion.  Assuming that all
supernovae are independently drawn from the same luminosity function with unknown dispersion, the total covariance matrix is
$\mathbf{C}=\mathbf{C}_{m}+\sigma_I^2\mathbf{I}$ making $\mathbf{C}$ a function of the model parameter $\sigma_I^2$.  I take the intrinsic scatter
to be parameterized by
the variance, not the standard deviation.

The determination of the best-fit and confidence regions for the parameters follows the standard procedure of minimizing
and mapping isocontours on the surface of Eqn.~\ref{likelihood:eqn}.

The prevalent supernova-cosmology analysis proceeds differently.  The intrinsic supernova variance is not treated as a fit parameter: the first term in
 the log-likelihood in Eqn.~\ref{likelihood:eqn} is ignored, some initial guess of $\sigma_I^2$ is included in $\mathbf{C}$, and
the $\chi^2$ (the second term in Equation~\ref{likelihood:eqn}) is minimized to get the best-fit parameters.  Then holding those parameters fixed,  the value of $\sigma_I^2$ that gives
 $\chi^2/dof=1$ is determined.  This process is repeated until the fits converge to stationary values.  Alternatively, this process is applied to a low-redshift
 subsample from which a $\sigma_I^2$ is measured as the dispersion from the linear Hubble law, and is inserted in the data
 covariance matrix of the full sample.  The closeness of the resulting $\chi^2/dof$ to unity
 checks the consistency between the dispersion of the training and full sets.

\section{Simulation}
\label{simulation:sec}
I simulate experiments specified by the number of supernovae they produce, either $N=50$ or 1000 uniformly distributed from
$0.08 \le z \le 0.8$, and the distance modulus measurement uncertainty per supernova $\sigma_s$,
either 0.05, 0.1, 0.2, or  $0.02+0.025z$ mag.  The data are
supplemented with an additional 100 SNe at $z=0.05$ each with a measurement uncertainty of 0.02 mag.
The measurement covariance is $C_{m,ij} = \delta_{ij} \sigma_s^2$.
The supernovae have an intrinsic dispersion of $\sigma_I=0.1$ mag.  The set of experimental realizations for each case is generated with the same
random-number generator seed.
All experiments occur in a flat $\Lambda$CDM universe with $\Omega_M=0.27$ and $w=-1$.

The data from each realized experiment are analyzed in two ways.
First, the data are fit using the full Equation~\ref{likelihood:eqn} to a model with a flat-universe dark-energy cosmology with
constant equation of state parameterized by $\Omega_M$ and  $w$,
and an intrinsic supernova dispersion $\sigma_I^2$.  This is referred to as the $\ln{L}$ fit.
Second, the data are initially fit to the cosmological model but holding $\sigma_{i}^2=0$ fixed.  Then, holding the best-fit
cosmological parameters fixed, the value of $\sigma_{i}^2$ that gives $\chi^2/dof=1$ is calculated.  This process is repeated twice more
starting with the updated values of $\sigma^2_I$; I find that the fit results converge after three iterations.    These are referred to as the $\chi^2$ fits.

For each type of experiment, I generate an ensemble of realizations each analyzed using the $\ln{L}$ and iterated $\chi^2$ fits.  
For the $N=50$ experiments I generate 5000 realizations  and for the $N=1000$ experiments, 1000 realizations.
The fitting is performed with the MIGRAD minimization
of the Minuit \citep{James:1975dr} implementation in ROOT \citep{Antcheva:2009zz}.  The parameter confidence intervals are taken directly
from the extrema of the $\mathcal{L}_{min}+1$ contours 
(using the MINOS function call); the contours can be asymmetric around the
extrema so my quoted uncertainties are half the interval length.

\begin{figure}
\epsscale{.8}
\plotone{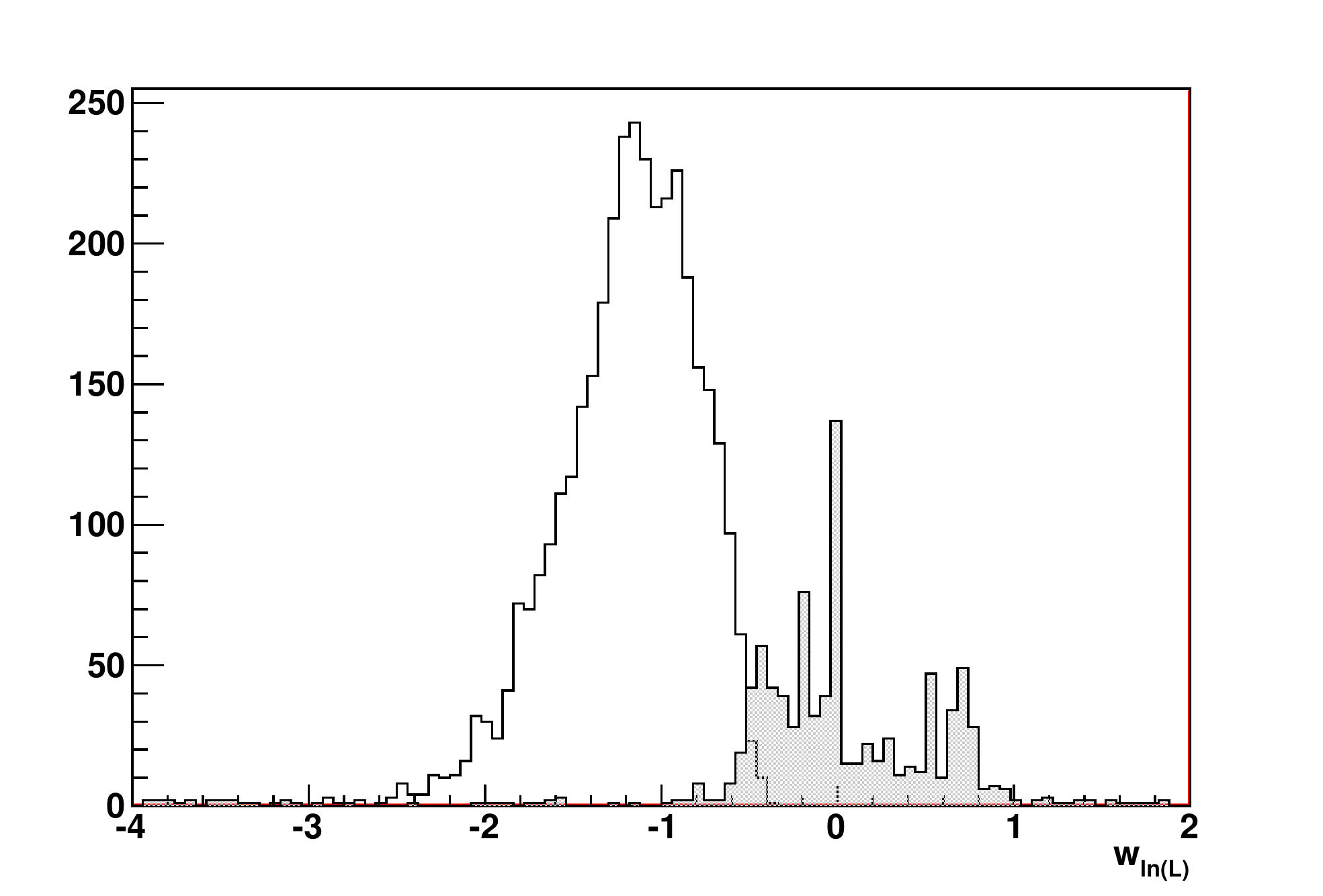}
\caption{The histogram of the best-fit $w$ from the $\ln{L}$ fit of the $N=50$, $\sigma_I=0.2$
run, the one with largest parameter uncertainties among the cases considered in this study.
The open curve includes  fits that succeed in determining the uncertainties in $\Omega_M$, the shaded curve includes those that fail.  The fitter often fails when the solution converges toward
 $w=0$.\label{failed:fig}}
\end{figure}

In terms of the fit, the cosmology model is pathological when $w=0$ and the dark energy is dynamically indistinguishable from non-relativisitic matter.
The minimization can fail if the maximum likelihood approaches $w=0$ and the parameter-uncertainty determination may fail when $w=0$ falls within
the accepted confidence region.  Given that the input cosmology has $w=-1$, the fitter encounters this condition with regularity only when the data quality
is poor.  This is seen in
Figure~\ref{failed:fig}, which shows the histogram of the best-fit $w$ from the $\ln{L}$ fit of the $N=50$, $\sigma_I=0.2$
run.  The open curve represents fits that succeeded in getting the asymmetric uncertainties in $\Omega_M$, the shaded curve represents those that failed.
The other runs with more SNe and/or lower measurement uncertainty produce significantly fewer or no such failures.
In my analysis I include only those realizations with successful uncertainty determination; the exclusion of the failed fits is not expected to
bias distributions of parameter uncertainties nor the determination of $\sigma_I^2$.

The $\chi^2$ fits converge to stable values by the third iteration. For example in the $N=50$, $\sigma_s=0.2$ run,
the distribution of the shift in $\sigma_I^2$ between
the second and third iterations has a mean of $5.8\times10^{-8}$ and an RMS of $1.8\times10^{-7}$, both small compared to the input $\sigma_I^2=0.01$.

For each $N$--$\sigma_s$ pair
I calculate the averages of the cosmological-parameter uncertainties and the best-fit and uncertainties for the 
intrinsic-dispersion parameter.  To directly compare the two fitters, I also calculate the mean and RMS
of the difference in the uncertainties they return.
The results are given in Table~\ref{results:tab}.

\begin{table}
\scriptsize
\begin{center}
\caption{The averages of the cosmological-parameter uncertainties and the best-fit and uncertainties for the 
intrinsic-dispersion parameter for all the simulated data sets.  Also tabulated are the average and RMS of the
parameter uncertainties subtracted by those of the $\log{L}$ fits (denoted by $\Delta$).  The data are analyzed either with the  $\ln{\det \mathbf{C}}$
term in Eqn.~\ref{likelihood:eqn} (denoted by $\ln{L}$) or without and holding $\sigma_I^2$ fixed . For the latter case, results are given for one or three iterations (denoted
by $\chi^2_1$ and $\chi^2_3$ respectively). \label{results:tab}}
\begin{tabular}{c|cc|cccccccc}
\tableline\tableline
$N$&$\sigma_s$&Fit&$\langle\sigma(\Omega_M)\rangle$ &$\langle\Delta \sigma(\Omega_M)\rangle$&$RMS(\Delta \sigma(\Omega_M))$&$\langle\sigma(w)\rangle$&$\langle\Delta \sigma(w)\rangle$&$RMS(\Delta\sigma(w))$& $\langle \sigma_I^2 \rangle$ &$\langle\sigma(\sigma_I^2)\rangle$ \\
\tableline
\multirow{12}{*}{50}&\multirow{3}{*}{0.05}&$\ln{L}$&0.191& $\dots$ & $\ldots$&0.311& $\dots$ & $\ldots$&0.00990&0.00126\\
&&$\chi^2_1$&0.080&-0.11794&   0.1506&0.091&-0.22015&   0.0337&0.01008 & $\ldots$\\
&&$\chi^2_3$&0.205& 0.00101&   0.0019&0.312& 0.00129&   0.0018&0.01003 & $\ldots$\\
\cline{2-11}
&\multirow{3}{*}{0.1}&$\ln{L}$&0.243& $\dots$ & $\ldots$&0.360& $\dots$ & $\ldots$&0.00992&0.00137\\
&&$\chi^2_1$&0.127&-0.11839&   0.2486&0.120&-0.23983&   0.0357&0.01011 & $\ldots$\\
&&$\chi^2_3$&0.260& 0.00121&   0.0094&0.360& 0.00101&   0.0050&0.01005 & $\ldots$\\
\cline{2-11}
&\multirow{3}{*}{0.2}&$\ln{L}$&0.267& $\dots$ & $\ldots$&0.458& $\dots$ & $\ldots$&0.00985&0.00144\\
&&$\chi^2_1$&0.179&-0.14510&   0.2027&0.186&-0.26416&   0.0366&0.01012 & $\ldots$\\
&&$\chi^2_3$&0.332& 0.00264&   0.0386&0.452& 0.00105&   0.0118&0.01009 & $\ldots$\\
\cline{2-11}
&\multirow{3}{*}{slope}&$\ln{L}$&0.266& $\dots$ & $\ldots$&0.396& $\dots$ & $\ldots$&0.00988&0.00137\\
&&$\chi^2_1$&0.157&-0.12438&   0.2123&0.162&-0.23292&   0.0397&0.01012 & $\ldots$\\
&&$\chi^2_3$&0.283& 0.00107&   0.0145&0.395& 0.00095&   0.0062&0.01006 & $\ldots$\\
\tableline
\multirow{12}{*}{1000}&\multirow{3}{*}{0.05}&$\ln{L}$&0.040& $\dots$ & $\ldots$&0.087& $\dots$ & $\ldots$&0.01000&0.00052\\
&&$\chi^2_1$&0.017&-0.02280&   0.0035&0.037&-0.04995&   0.0032&0.01003 & $\ldots$\\
&&$\chi^2_3$&0.040& 0.00003&   0.0000&0.087& 0.00006&   0.0001&0.01002 & $\ldots$\\
\cline{2-11}
&\multirow{3}{*}{0.1}&$\ln{L}$&0.050& $\dots$ & $\ldots$&0.109& $\dots$ & $\ldots$&0.00999&0.00076\\
&&$\chi^2_1$&0.031&-0.01944&   0.0095&0.062&-0.04632&   0.0059&0.01008 & $\ldots$\\
&&$\chi^2_3$&0.050& 0.00004&   0.0003&0.109& 0.00009&   0.0005&0.01003 & $\ldots$\\
\cline{2-11}
&\multirow{3}{*}{0.2}&$\ln{L}$&0.081& $\dots$ & $\ldots$&0.167& $\dots$ & $\ldots$&0.00998&0.00123\\
&&$\chi^2_1$&0.074&-0.00670&   0.3741&0.088&-0.07885&   0.0114&0.01023 & $\ldots$\\
&&$\chi^2_3$&0.081& 0.00008&   0.0013&0.167& 0.00016&   0.0024&0.01011 & $\ldots$\\
\cline{2-11}
&\multirow{3}{*}{slope}&$\ln{L}$&0.057& $\dots$ & $\ldots$&0.116& $\dots$ & $\ldots$&0.01000&0.00076\\
&&$\chi^2_1$&0.035&-0.02110&   0.0096&0.065&-0.05052&   0.0063&0.01007 & $\ldots$\\
&&$\chi^2_3$&0.057& 0.00003&   0.0005&0.116& 0.00007&   0.0011&0.01004 & $\ldots$\\
\tableline
\end{tabular}
\end{center}
\end{table}

The distributions are skewed by amounts that depend on the statistic and the experimental configuration.  It is well known that parameter confidence regions are not elliptical and that the
size of the region depends on where the best fit falls \citep{2002PhRvD..65j3512W}.  In addition, as seen in Figure~\ref{failed:fig}, the fitter fails preferentially in the tail of the distribution where
$w=0$ is favored.  The asymmetries are therefore accentuated when
parameter uncertainties are large; among the cases considered in this paper the $\Omega_M$ fits of the $N=50$ runs and the $w$
fit of the $N=50$, $\sigma_s=0.2$ run are particularly affected.  In these extreme cases, the quoted averages should be interpreted with care.

\begin{figure}
\plotone{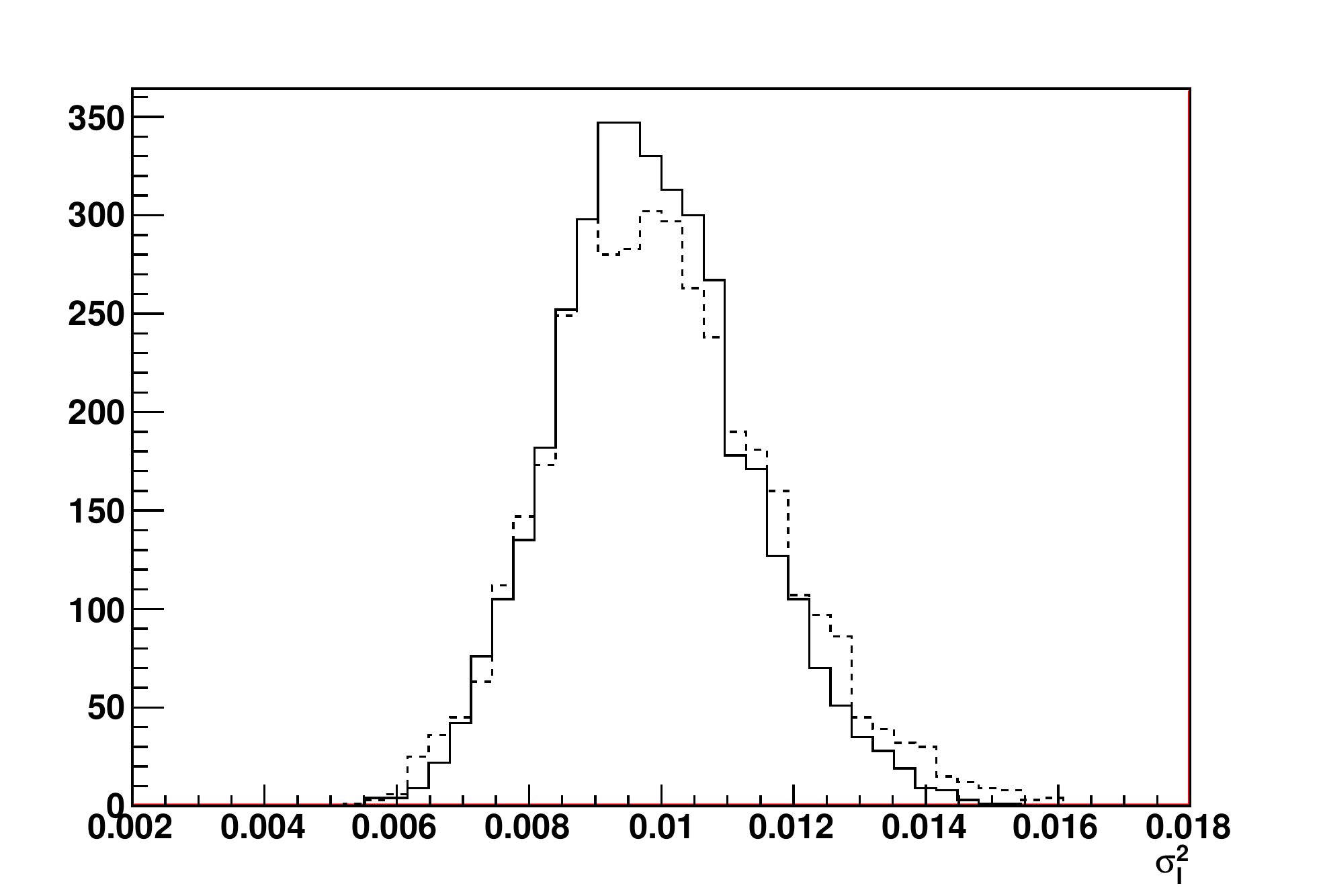}
\caption{The histograms of $\sigma^2_I$ as determined by the $\ln{L}$ (solid) and $\chi^2$ (dashed) fits for $N=50$ and $\sigma_s=0.2$.
The distributions are slightly asymmetric with broader tails at larger values.  The $\ln{L}$ fits fail more frequently than $\chi^2$ fits, for
direct comparison both histograms include realizations that succeed in both fits.
\label{intrinsicDispersionHist:fig}}
\end{figure}

\begin{figure}
\epsscale{.8}
\plotone{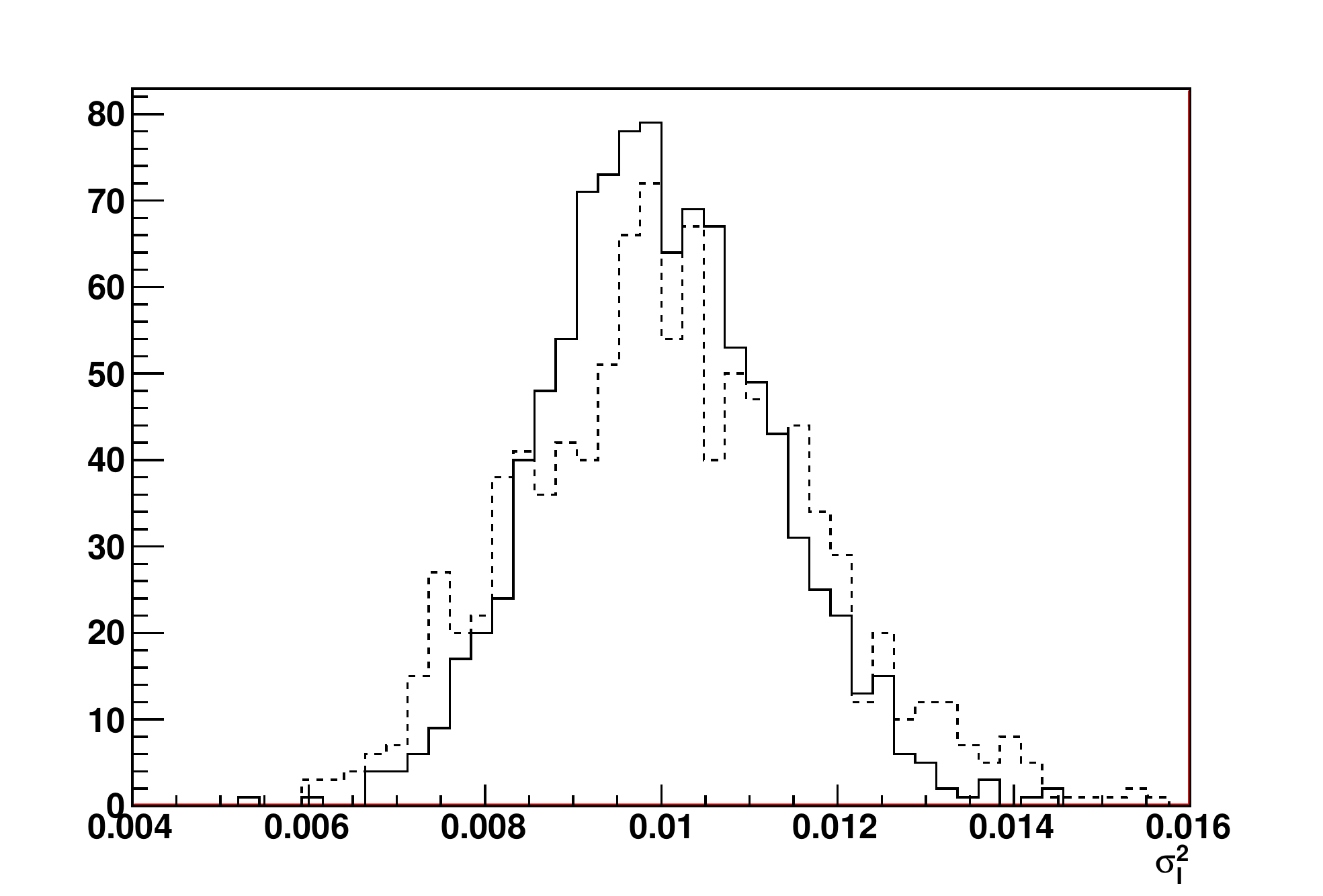}
\plotone{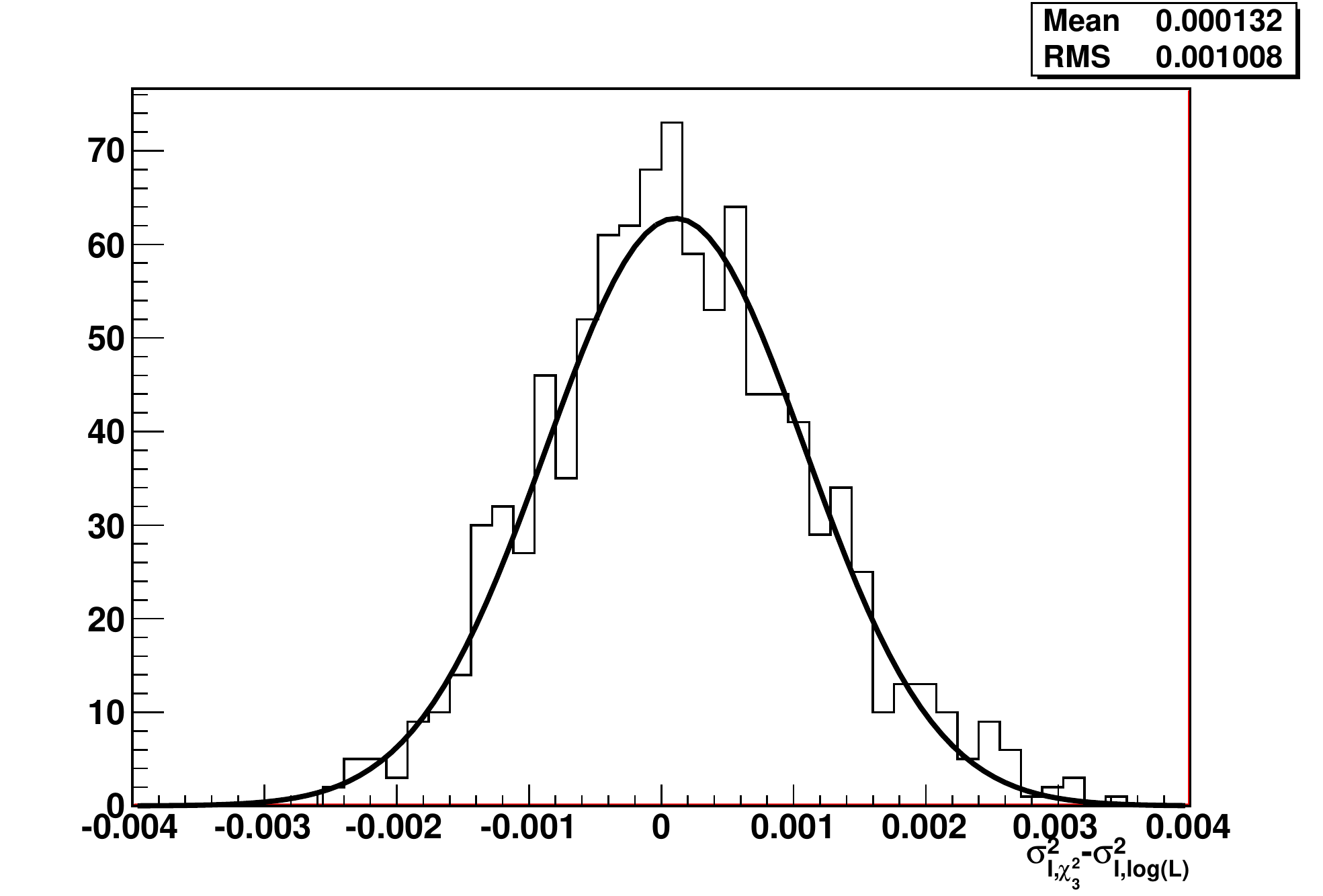}
\caption{The top plot shows histograms of $\sigma^2_I$ as determined by the $\ln{L}$ (solid) and $\chi^2$ (dashed) fits for $N=1000$ and $\sigma_s=0.2$.
Unlike the $N=50$, $\sigma_s=0.2$ case, both fitting methods succeed for all realizations.
The bottom plot shows the histogram of their difference $\sigma^2_{i,\chi^2}-\sigma^2_{i,\ln{L}}$ with the best-fit Gaussian overplotted. Note that the
distribution is slightly asymmetric with all the points on the right-side tail falling well above the Gaussian curve.  \label{intrinsicDispersionDiff:fig}}
\end{figure}

Both the $\ln{L}$ and $\chi^2$ fits return asymmetric $\sigma^2_i$ distributions.  The asymmetry is more pronounced when the $\ln{L}$ fits
have higher $\sigma(\sigma^2_I)$  and the corresponding $\chi^2$-fit distributions are even more skewed.
For larger $N$ and/or as $\sigma_s$ decreases the averages of the $\ln{L}$-fit distribution approach the input intrinsic dispersion.
The averages of the $\chi^2$-fit distribution also approach the input as $\sigma_s$ decreases but the  bias remains when going from $N=50$ to
$N=1000$.  To illustrate,  Figure~\ref{intrinsicDispersionHist:fig} plots for the $N=50$, $\sigma_s=0.2$ run histograms of $\sigma_I^2$
from both fits. The two distributions are different and while both are
asymmetric that of the the $\chi^2$-fit  has a broader tail.  Table~\ref{results:tab} gives $\langle \sigma_{I,\ln{L}}^2 \rangle=0.00985$ 
and $\langle \sigma_{I,\chi^2}^2 \rangle=0.01009$.
Figure~\ref{intrinsicDispersionDiff:fig} shows the corresponding histograms  for the $N=1000$, $\sigma_s=0.2$ run, and  a histogram
of the differences in the intrinsic dispersions from the two fits,
$\sigma^2_{I,\chi^2}-\sigma^2_{I,\ln{L}}$.  Here the asymmetry is subtle and is more clearly seen in the differences,
and  $\langle \sigma_{I,\ln{L}}^2 \rangle=0.00998$ approaches the input 0.001
whereas $\langle \sigma_{I,\chi^2}^2 \rangle= 0.01011$ remains offset.

Though not directly applicable to the cases simulated for this paper,
I point out that when all supernovae have the same measurement uncertainty the distribution of intrinsic dispersions that give $\chi^2/dof=1$ is known trivially.
An experiment with a realized $\chi^2_R$ for an input intrinsic
dispersion $\sigma^2_{I0}$ has an inferred intrinsic dispersion
$$\sigma^2_I=\frac{(\sigma_s^2+\sigma^2_{I0})}{dof}\chi^2_R-\sigma_s^2.$$
The $\sigma^2_I$ distribution thus corresponds directly with the $\chi^2$ distribution.  This does not apply to the
experiments simulated in this paper that have a low-redshift
set of supernovae with measurement uncertainty that differs from those of the high-redshift set.

\begin{figure}
\epsscale{.8}
\plotone{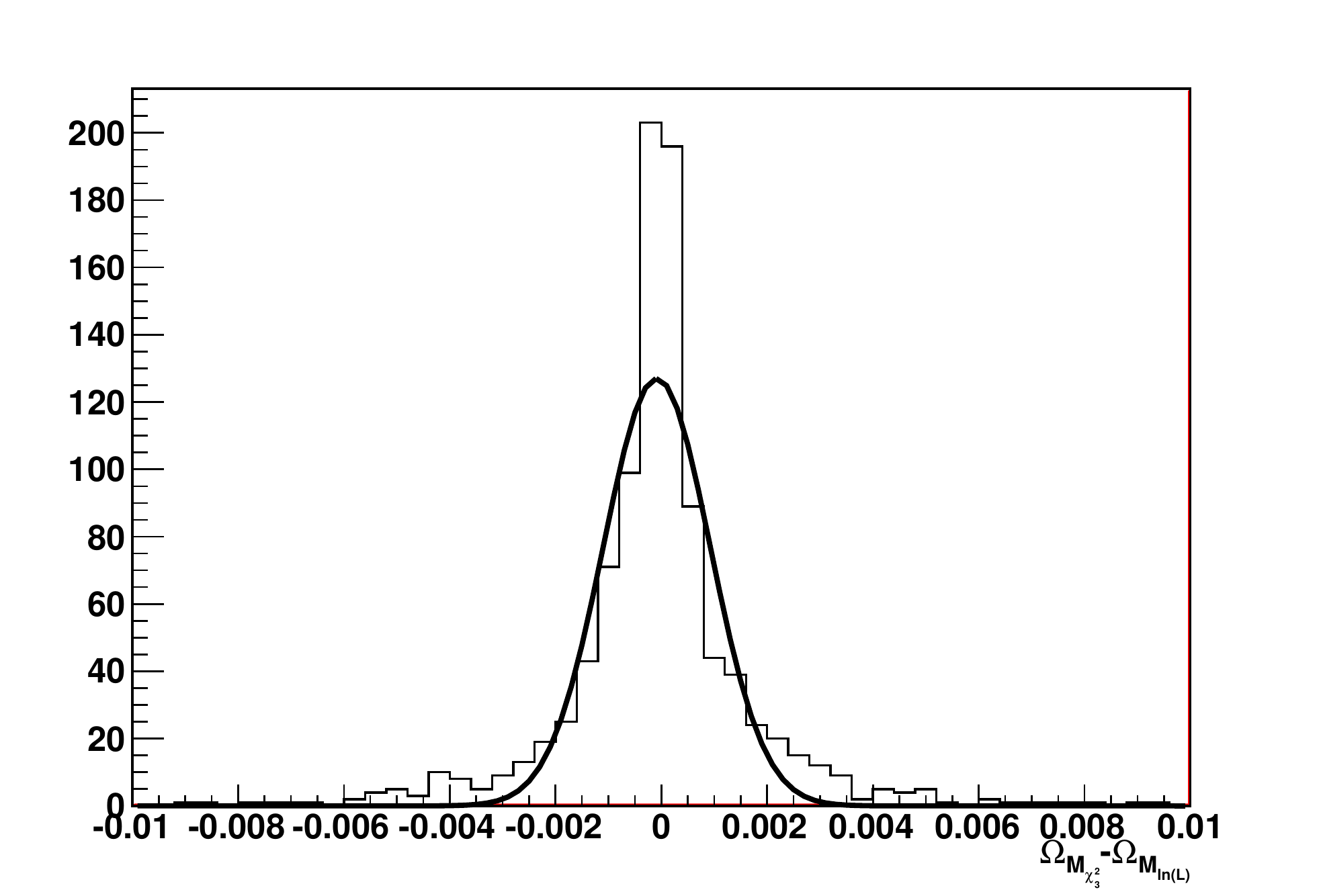}
\plotone{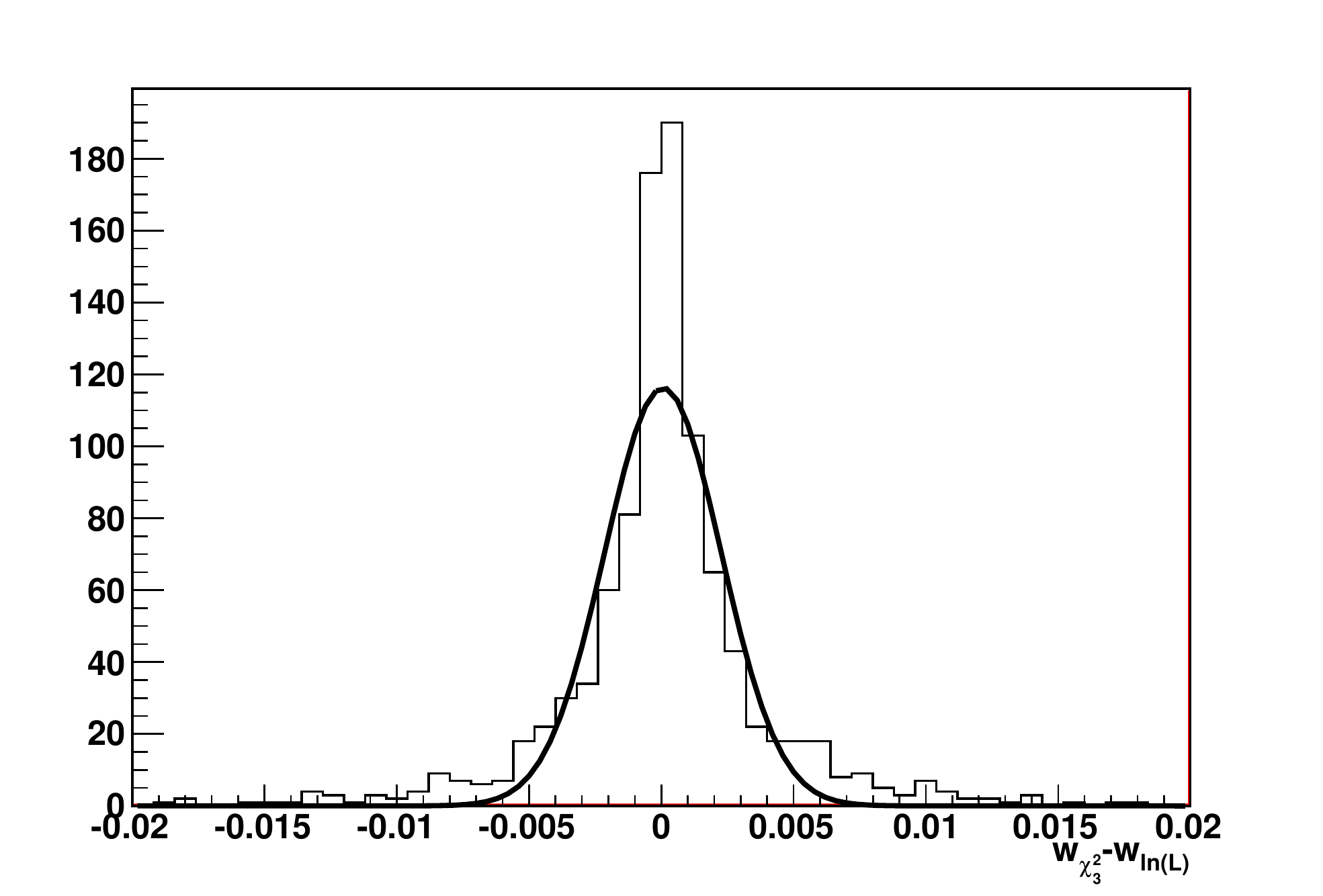}
\caption{Histograms of the difference in the best-fit parameters returned by the $\chi^2$ and $\ln{L}$ fitters, $\Omega_{M,\chi^2}-\Omega_{M,\ln{L}}$ (top) and $w_{\chi^2}-w_{\ln{L}}$ (bottom), for the
$N=1000$ and $\sigma_s=0.02$
experiment. Overplotted on each are the Gaussian best-fits to the data. \label{residuals0:fig}}
\end{figure}

The best-fit cosmological parameters differ for the two fitters only when they deduce different intrinsic dispersions.  As  seen
in Table~\ref{results:tab} and the example shown in Figure~\ref{intrinsicDispersionDiff:fig}, the $\sigma^2_I$'s returned by the two fits
agree with little bias within expected measurement uncertainties.  I confirm that the fits also find similar optimal $\Omega_M$ and $w$.
Figure~\ref{residuals0:fig} shows histograms for $\Omega_{M,\chi^2}-\Omega_{M,\ln{L}}$ and $w_{\chi^2}-w_{\ln{L}}$ from the representative $N=1000$ and $\sigma_s=0.2$ run.
The both are highly peaked around zero with ranges much smaller than the statistical measurement uncertainty.

\begin{figure}
\epsscale{.8}
\plotone{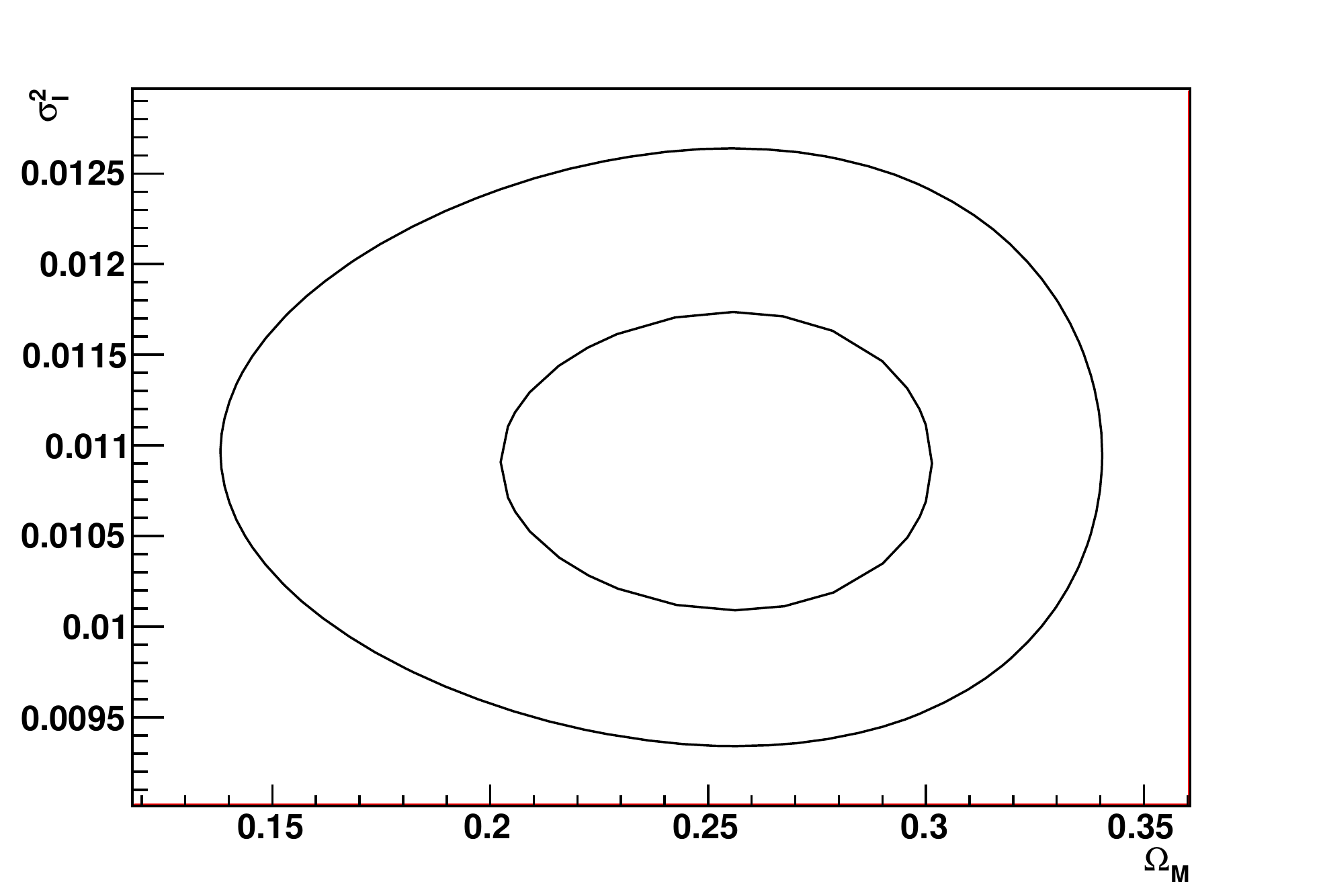}
\plotone{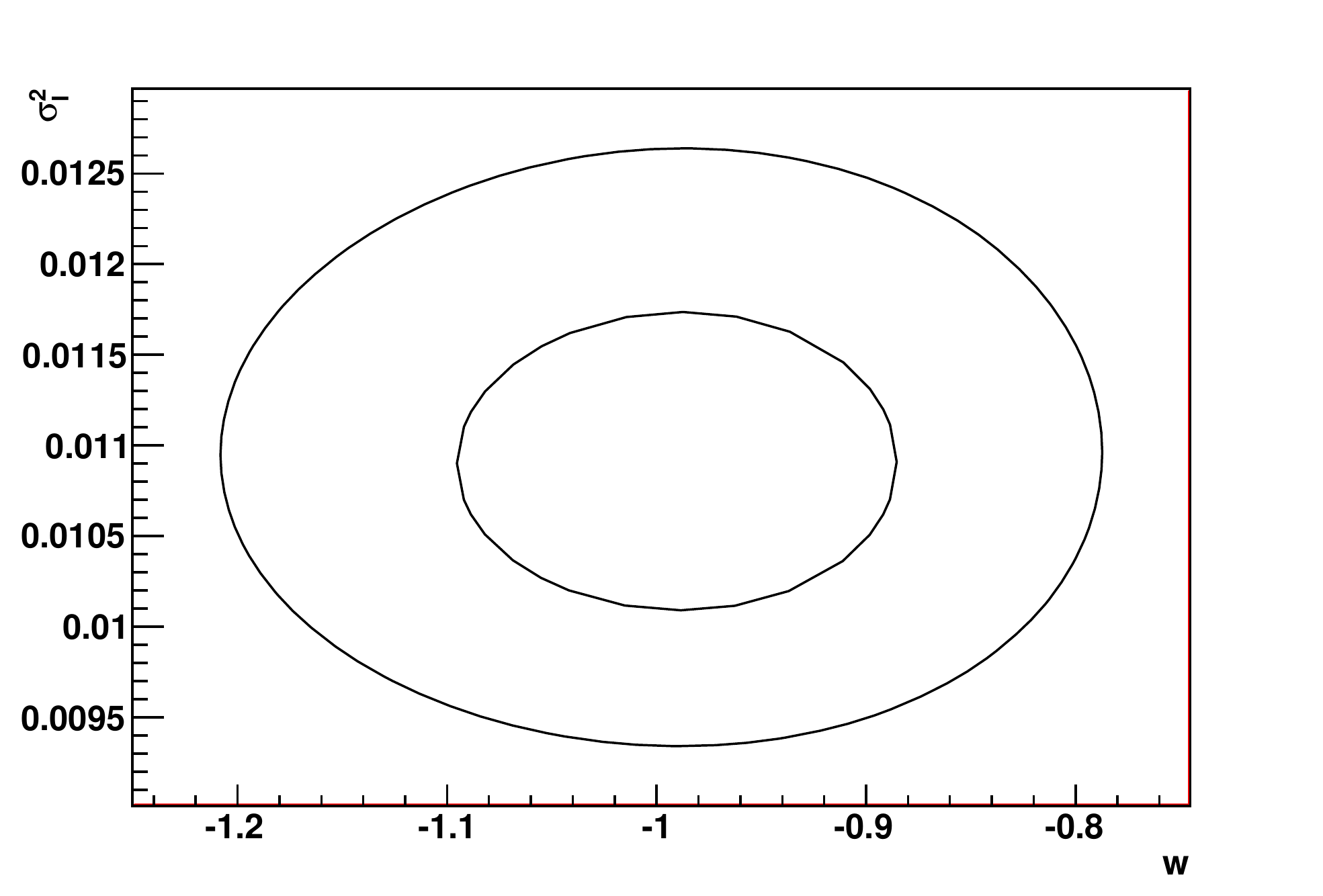}
\caption{The $\Omega_M$--$\sigma^2_I$ (top) and $w$--$\sigma^2_I$ (bottom) 1- and 2-$\sigma$ confidence regions  for one realization of a $N=1000$ and $\sigma_s=0.1$
experiment. The parameters
have small 0.019 correlation. \label{contours:fig}}
\end{figure}

In an individual realization of an experiment
the covariance between the intrinsic dispersion and the cosmological
parameters in the $\ln{L}$ fits can be non-zero.  This is illustrated by a typical realization of a $N=1000$ and  $\sigma_s=0.1$ experiment;
the correlation coefficients between $\sigma^2_I$ and $\Omega_M$ and $w$ are $-0.018$ and $ 0.019$ respectively. The 
corresponding 1- and 2-$\sigma$ confidence regions in $\Omega$--$\sigma^2_I$ and $w$--$\sigma^2_I$ space  are shown in Figure~\ref{contours:fig}.
The likelihood is pronouncedly asymmetric for $\Omega_M$ and $w$, whereas it
is close to symmetric in $\sigma^2_I$. 
The $\chi^2$ fits do not provide an uncertainty for the intrinsic dispersion nor their propagated effect on the other parameters.

The average parameter uncertainties for a given run must differ between fitters.
The minimum of $\mathcal{L}$ in the  $\ln{L}$ fit is less than (or equal to) the minimum $\chi^2$  so
$\mathcal{L}^{min}_{\ln{L}}+1 \le \mathcal{L}^{min}_{\chi^2}+1$, the conditions
that define the 1-$\sigma$ contours.  Also, the extra $\sigma^2_I$ dimension in the $\ln{L}$ fit opens room
for  a broader range of acceptable cosmological-parameter values to be contained within the confidence region.

\begin{figure}
\epsscale{.8}
\plotone{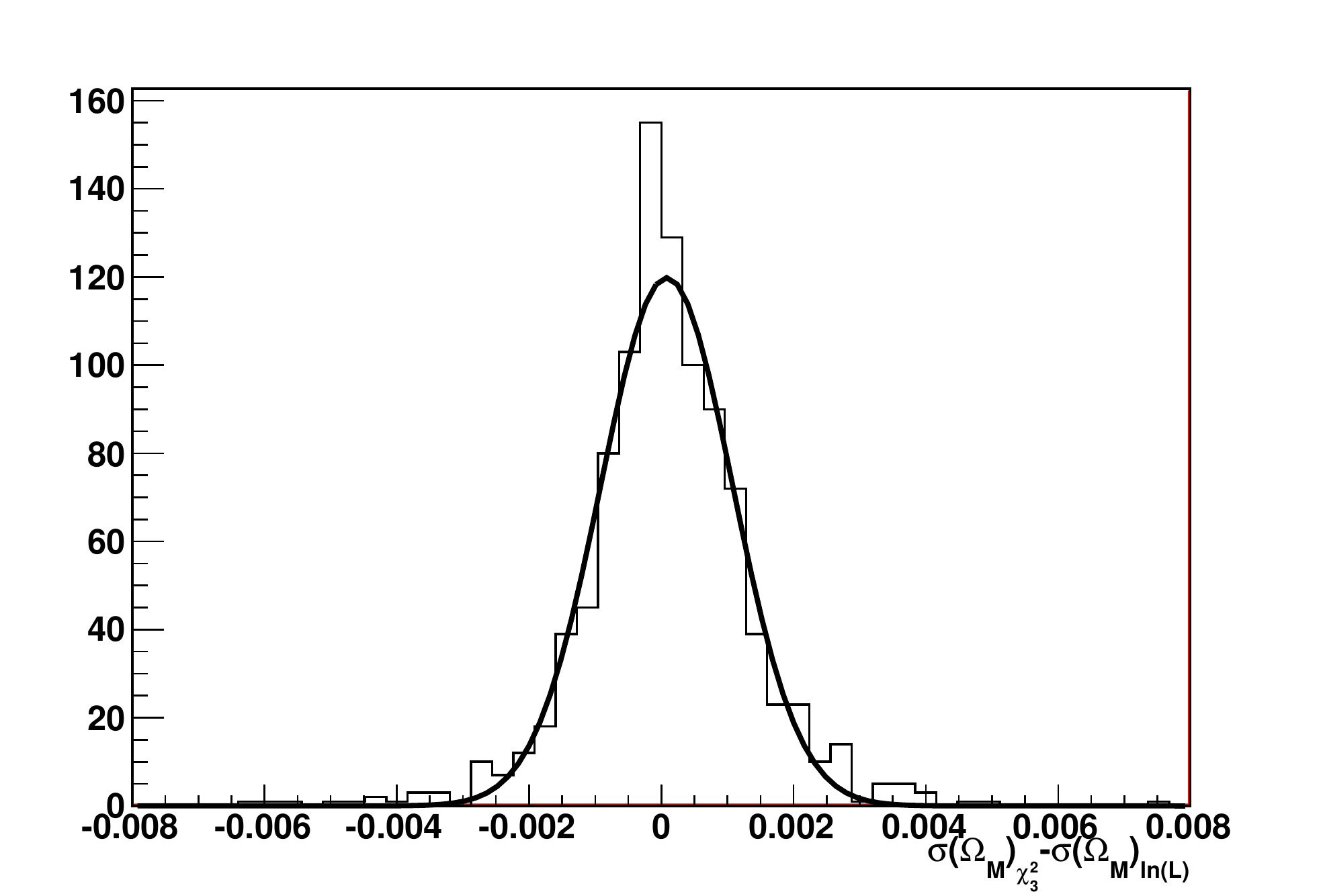}
\plotone{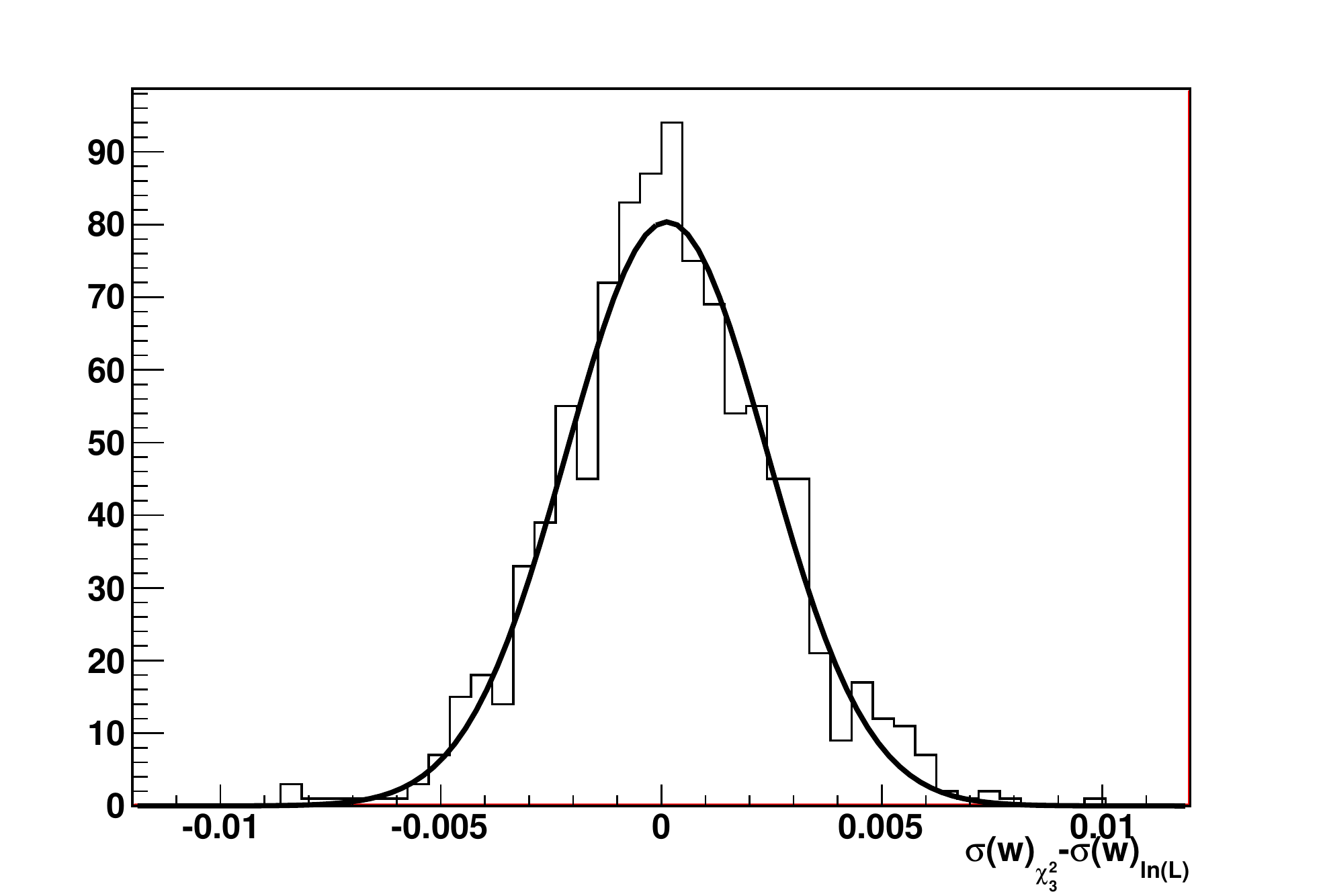}
\caption{Histograms of the difference in the parameter uncertainties returned by the $\chi^2$ and $\ln{L}$ fitters, $\sigma(\Omega_M)_{\chi^2}-\sigma(\Omega_M)_{\ln{L}}$ (top)
and $\sigma(w)_{\chi^2}-\sigma(w)_{\ln{L}}$ (bottom), for the
$N=1000$ and $\sigma_s=0.02$
experiment. Overplotted on each are the Gaussian best-fits to the data. \label{residuals:fig}}
\end{figure}

Except for the $\Omega_M$ uncertainties in the $N=50$ runs, both fits give similar average 
uncertainties in the cosmological parameters.  On a per-realization level, the average and RMS of the difference  between the $\chi^2$ and
$\ln{L}$ fits ($\langle\Delta \sigma(\Omega_M)\rangle$, $RMS(\Delta \sigma(\Omega_M))$,
 $\langle\Delta \sigma(w)\rangle$, and $RMS(\Delta \sigma(w))$in Table~\ref{results:tab}) are small
compared to the uncertainties themselves.
The distributions of  the difference in parameter uncertainties between the third $\chi^2$-fit iteration
and the $\ln{L}$ fit $\sigma(w_a)$ are shown in Figure~\ref{residuals:fig} for the case of $N=1000$ and $\sigma_s=0.02$.
Although they both are close to Gaussian, 
there is a slight excess in the high end (corresponding to larger $\chi^2$- or smaller $\ln{L}$-fit uncertainties) just as is the case in
the distribution of $\sigma^2_I$ differences shown in Figure~\ref{intrinsicDispersionDiff:fig}.

\citet{2010ApJ...716..712A} have shown that supernova samples from different observatories exhibit different intrinsic magnitude dispersions.
I run the $N=1000$, $\sigma_s=0.1$ case fitting for two
intrinsic dispersion parameters, one for the nearby $z=0.05$ set and another for the higher-redshift set. 
The fitters return averages for the intrinsic dispersion uncertainties,  $\langle\sigma(\sigma_I^2)\rangle$, of 0.00148 for low redshift and 0.00090 for high redshift.   These
numbers provide a quantitative measure of how well possible systematic differences between the two populations could be resolved.
For comparison,  $\langle\sigma(\sigma_I^2)\rangle=0.00076$ when a single $\sigma^2_I$ is fit for all supernovae.

\section{Conclusions}
\label{conclusions:sec}
I have shown how to fit for cosmological parameters with SNe Ia when the intrinsic dispersion of the standard candle is unknown.
My standard likelihood function has not been used in cosmological analysis to date.  I show via simulation
that, on average, our likelihood function is maximal at the values of the input parameters including the intrinsic dispersion.  The presented
and previously used iterative fitting methods do not
give biases in the best-fit cosmological parameters and any differences in a single experiment are due to realization scatter.
The fitter methods  do return different intrinsic dispersions and parameter uncertainties.
The procedure presented here has the advantage
that it includes the covariance of the intrinsic dispersion with the other parameters in its error propagation, and the fit is done in a single iteration.

The methodology can be extended to cases where multiple dispersion parameters are fit.  I present an example
taking the low- and high-redshift sets as being drawn from different magnitude distributions.
The same approach can be used to check whether different supernova subsets
(tagged for example by redshift, host-galaxy characteristics or spectral features) exhibit statistically significant
differences in their population characteristics.

The approach is appropriate for any analysis that uses a statistic for which the tracer has an intrinsic dispersion that must be determined from the data.
For example, in weak gravitational
lensing the measurement of correlated shear is obscured by the unknown intrinsic shape of individual galaxies.
The intrinsic dispersion in galaxy ellipticities can be made a fit parameter determined simultaneously with those of cosmological interest.

Inclusion of the likelihood-function normalization when fitting is not new to astronomy nor cosmology;
\citet{1995ApJ...438..322W} showed its importance in shot-noise-dominated photometry
and it is retained in other cosmological analyses \cite[see e.g.][]{2002MNRAS.335.1193B,
2010MNRAS.tmp.1232T}. \citet{2010arXiv1009.5443H} do include a fit parameter in the data covariance for their supernova analysis although there it
serves as a hyperparameter of the Gaussian-process prior on $w(z)$. 
\citet{2010ApJ...717...40K} include the normalization term; though containing no fit parameters it is needed
to directly compare the $\chi^2$'s derived from different light-curve models.

This paper gives a simplified view of how the standard candle nature of SNe Ia is used in cosmology analysis.  SNe Ia are in fact calibrated
candles; independent observables (light-curve shape, colors, spectral features)
are correlated with peak absolute magnitude to correct and lower the dispersion in distance determinations.
I advocate that intrinsic dispersion be measured as a fit parameter from the data
simultaneously with the magnitude-correction and cosmological parameters.  This provides a new perspective
in how we search for magnitude corrections that make SNe Ia better
calibrated candles.  In the past we have sought parameterized magnitude corrections that minimize distance dispersion;
we can now seek corrections and their inferred intrinsic dispersions that are most consistent with observations and are statistically favored over having
no correction.
Application of this technique to real SN data sets is the subject of ongoing work.

\acknowledgments
I acknowledge fruitful discussions with Eric Linder, David Rubin, and Ramon Miquel.
This work was supported by the Director, Office of Science, Office of High Energy Physics, 
of the U.S.\ Department of Energy under Contract No. DE-AC02-05CH11231.

\bibliography{/Users/akim/Documents/alex}

\end{document}